# Electronic Liquid Crystal State in the High-Temperature Superconductor $YBa_2Cu_3O_{6.45}$

V. Hinkov,[1*] D. Haug,[1] B. Fauqué,[2] P. Bourges,[2] Y. Sidis,[2] A. Ivanov,[3] C. Bernhard,[4] C. T. Lin,[1] B. Keimer[1]

[1]Max-Planck-Institut für Festkörperforschung, Heisenbergstr. 1, D-70569 Stuttgart, Germany
[2]Laboratoire Léon Brillouin, CEA-CNRS, CE-Saclay, F-91191 Gif-sur-Yvette, France
[3]Institut Laue-Langevin, 6 Rue J. Horowitz, F-38042 Grenoble cedex 9, France
[4]Department of Physics and FriMat Center for Nanomaterials, University of Fribourg, Chemin du Musée 3, CH-1700 Fribourg, Switzerland

*To whom correspondence should be addressed; E-mail: v.hinkov@fkf.mpg.de.

**Electronic phases with symmetry properties matching those of conventional liquid crystals have recently been discovered in transport experiments on semiconductor heterostructures and metal oxides at milli-Kelvin temperatures. We report the spontaneous onset of a one-dimensional, incommensurate modulation of the spin system in the high-temperature superconductor $YBa_2Cu_3O_{6.45}$ upon cooling below ~150 K, while static magnetic order is absent above 2 K. The evolution of this modulation with temperature and doping parallels that of the in-plane anisotropy of the resistivity, indicating an electronic nematic phase that is stable over a wide temperature range. The results suggest that soft spin fluctuations are a microscopic route towards electronic liquid crystals, and nematic order can coexist with high-temperature superconductivity in underdoped cuprates.**

The electronic states near the Fermi level of high-temperature superconductors derive from the hybridized *d*- and *p*-orbitals of copper and oxygen ions in a square-planar network. At a doping level of 1/8 hole per Cu ion, experimental work on a specific superconducting cuprate family, $(La,Nd)_{2-x}(Sr,Ba)_xCuO_4$ (La214), has shown that the two-dimensional electron system in the $CuO_2$ layers can support a state with uniaxial spin (*1–3*) and charge (*1,4,5*) order ("stripes"). Static stripe order implies that both translational and rotational symmetries of the copper-oxide square lattice are spontaneously broken. More unusual "electronic liquid crystal" states (*6*) that break the rotational symmetry of the lattice while at least partially preserving its translational symmetry can arise from quantum fluctuations of stripes(*6–8*), or from Fermi surface instabilities (*9–11*). Electronic nematic states have recently been discovered in semiconductor heterostructures (*12*) and in the bulk transition metal oxide $Sr_3Ru_2O_7$ (*13*). In both cases, however, they are stable only at milli-Kelvin temperatures and in high magnetic fields, and have thus far only been probed by transport measurements. We use neutron scattering to address the role of magnetic degrees of freedom in driving the formation of electronic liquid crystals, and to explore the presence of liquid-crystalline order in the cuprates.

In the prior experiments on semiconductor heterostructures and on $Sr_3Ru_2O_7$, the nematic director was aligned by external magnetic fields, resulting in a strong macroscopic anisotropy of the current flow (*12,13*), analogous to the alignment of nematic domains in conventional liquid crystals by electric fields or confining walls. In the cuprates, subtle crystallographic distortions can serve as aligning fields for symmetry-broken electronic phases. They reduce the fourfold rotational symmetry of the $CuO_2$ layer to a twofold rotational or mirror symmetry by introducing a slight (~1%) difference between the in-plane lattice parameters. In stripe-ordered La214, the stripe domains in every $CuO_2$ layer are aligned by such a twofold axis, but the layers are stacked in such a way that the global crystal symmetry is tetragonal, and no macroscopic anisotropy is observed (*1*). The situation is more favorable in metallic $YBa_2Cu_3O_{6+x}$ (Y123), where a macroscopic orthorhombic crystal structure is stabilized by one-dimensional CuO chains lying between the $CuO_2$ bilayers (see the SOM for Supporting Text (*14*)). Indeed, an unexpectedly large, temperature and doping dependent in-plane anisotropy of the resistivity has been reported for this material (*15*). While this constitutes possible evidence of nematic order (*7,15*), clear signatures of an isotropic-to-nematic transition are not apparent in the transport data.

A determination of the spin-spin correlation function by magnetic neutron scattering has the potential to build a more robust case for nematic order, and to elucidate the underlying microscopic mechanisms. Our sample is a large array of untwined $YBa_2Cu_3O_{6.45}$ single crystals with superconducting transition temperature $T_c = 35$ K (*14,16*). In this work we show only magnetic excitations that are odd under the exchange of the two layers within a $CuO_2$ bilayer (*14,17*). Figures 1A–C provide an overview of the magnetic spectrum along the two perpendicular in-plane axes $a^*$ and $b^*$ at a temperature of 5 K. As previously observed for Y123 samples at similar doping levels (*17–20*), the magnetic intensity is concentrated around the in-plane wave vector $\mathbf{Q}_{AF} = (\pi/a, \pi/b)$ (the propagation vector of the antiferromagnetic state in undoped Y123), and the spectrum is nearly gapless (Fig. 3C). The new aspect uncovered by the experiments on twin-free samples presented here is the in-plane anisotropy of the spectrum, which exhibits an unusual evolution with energy and temperature. Whereas at high excitation energies, $E$, the spectrum is isotropic (Fig. 1C), maps of the magnetic spectral weight for $E < 15$ meV reveal a pronounced anisotropy that increases with decreasing energy (Fig. 1A, B). This anisotropy could not be recognized in prior work due to crystal twinning (Fig. 1D). Cuts through the spectrum with high instrumental resolution (Fig. 1E, F) reveal that the anisotropic intensity distribution is generated by two incommensurate peaks symmetrically displaced from $\mathbf{Q}_{AF}$ along $a^*$, while along $b^*$ the distribution is commensurate.

Figure 2 shows the temperature dependence of constant-energy cuts along the two high-symmetry axes. While the isotropic intensity distribution at high $E$ is temperature-independent, the anisotropy at low $E$ is strongly reduced with increasing temperature. Fits of the profiles to two Gaussians centered at $\mathbf{Q} = (\pi/a \pm \delta, \pi/b)$ yield excellent descriptions of the data at all temperatures (lines in Figs. 1E and 2). The results of the fits for E = 3 meV are plotted in Fig. 3. The intrinsic half-widths-at-half-maximum along $a^*$ and $b^*$, $\xi_a^{-1}$ and $\xi_b^{-1}$ (extracted from the Gaussian peaks after a correction for the instrumental resolution), are nearly identical and weakly temperature dependent. The zero-temperature offset suggests that magnetic long-range order is absent, reflecting the influence of magnetic quantum fluctuations, low-energy charge fluctuations and/or disorder.

The incommensurability $\delta$ exhibits an order-parameter-like behavior with an onset temperature

~ 150 K (Fig. 3A). Although $\delta$ can no longer be accurately determined (nor is it physically meaningful) once it becomes smaller than $\xi_a^{-1}$ (shaded area in Fig. 3A), this behavior indicates an underlying phase transition where $\delta \to 0$. At the same temperature, the **Q**-integrated spin susceptibility $\chi''(\omega)$ exhibits a strong upturn (Fig. 3B), which is also reminiscent of a broadened phase transition. A rounding of the singularities related to a phase transition is generally expected because the spin system is probed at nonzero energy, but additional contributions can arise from disorder and/or the orthorhombic distortion of the crystal structure.

Ordinary magnetic phase transitions are associated with the formation of static magnetic moments. We have performed zero-field muon-spin-relaxation (μSR) experiments at the πm3 beamline at the Paul-Scherrer-Institut in Villigen (CH), capable of routinely detecting static or slowly fluctuating electronic moments of order 0.01 $\mu_B$ per lattice site. Figure 4A shows that at temperatures above 10 K, the relaxation of muon spins implanted into our samples is entirely determined by nuclear spins. In agreement with prior work (*21*), an additional contribution to the μSR signal from low-energy electronic spin excitations is seen below 10 K. Manifestations of electronic magnetic moments that are static on the microsecond time scale probed by the muons are discernible only in the decay profiles below ~ 2 K.

Information about the spatial structure of these low-energy spin correlations was obtained by additional neutron scattering measurements with a spectrometer nominally set for zero energy transfer. Due to the finite energy resolution, spin fluctuations with energies below the instrumental resolution width of ~ 0.2 meV are probed in this "quasi-elastic" configuration. Representative data (Fig. 4B) show that the incommensurability of the quasi-elastic profile as well as its width $\xi_a^{-1}$ are nearly identical to those of the low-energy inelastic data plotted in Fig. 3A; $\xi_b^{-1}$ is somewhat smaller. The quasi-elastic intensity exhibits a significant upturn below 30 K. These data fit well into the "spin freezing" phenomenology of deeply underdoped cuprates (*20–24*). The characteristic frequency of electronic spin fluctuations is reduced with decreasing temperature; this is directly apparent in the inelastic neutron scattering data of Fig. 3C. Upon further cooling, the spin system gradually freezes into an ensemble of slowly fluctuating, finite-sized domains. The characteristic fluctuation rate of these domains progressively enters the frequency windows of quasi-elastic neutron scattering (*20,22,23*), μSR (*21,22*), and nuclear magnetic resonance (*24*) methods.

The results demonstrate that the spin correlations within the fluctuating domains are incommensurate. As the signal derives entirely from the $CuO_2$ planes (*14*), coupling between spins in the planes and the CuO chains can be ruled out as a substantial driving force of these correlations. Thus, our data demonstrate that the spin system in the $CuO_2$ planes of strongly underdoped Y123 becomes inherently unstable towards the formation of a uniaxial, slowly fluctuating spin texture at a critical temperature of ~ 150 K. The small (< 1%) orthorhombic distortion of the crystal structure serves as an aligning field for the incommensurate domains and leads to the large anisotropy of the neutron scattering pattern below 150 K. This scenario is also consistent with the observation of local uniaxial charge domains in the spin-glass state of other high temperature superconductors by scanning tunneling spectroscopy (*25*).

The relationship between the magnetic dynamics and the charge transport properties of $YBa_2Cu_3O_{6.45}$ can be discussed in the context of theoretical proposals for electronic nematic order. While in-plane anisotropies of both the spin fluctuation spectrum and the electrical resistivity in the absence of spin and charge order are generic features of nematic states in theoretical models of quasi-two-dimensional electron systems (*10,11*), nematic order has thus far been diagnosed solely on the basis of the spontaneous onset of anisotropic resistivity (*12,13*). In Y123, the case for nematicity is harder to make based on transport alone, because the aligning field is not tunable like the magnetic fields used in Refs. (*12,13*). The resistivity is thus anisotropic at all temperatures, and the putative isotropic-to-nematic transition is inevitably broadened. The pronounced enhancement of the in-plane resistivity ratio, $\rho_a/\rho_b$, below ~ 200 K (inset in Fig. 3B) is, however, suggestive of an underlying phase transition (*15*). The strikingly similar temperature dependence of the spectral weight of the anisotropic low-energy spin fluctuations (Fig. 3B) confirms this interpretation. In addition, both quantities exhibit a parallel evolution with energy and doping: Optical conductivity measurements on Y123 crystals with doping levels similar to ours indicate that the charge transport anisotropy is strongly reduced at excitation energies above ~ 20 meV (*26*), in good agreement with the crossover to isotropic spin fluctuations we observe. In more highly doped Y123 (where the orthorhombic distortion is enhanced), $\rho_a/\rho_b$ is reduced and much less temperature dependent (*15*), while the spectral weight of the low-energy collective spin excitations is strongly diminished (*17–19,27–29*). Although low-energy spin excitations are not entirely suppressed, this phenomenon has been termed a "spin-gap". We note that the dispersion and in-plane geometry of the excitations at energies exceeding the spin gap of $YBa_2Cu_3O_{6+x}$ ($x \geq 0.5$) above $T_c$ bear some resemblance to the low-energy excitations of $YBa_2Cu_3O_{6.45}$ (*19,28,29*). This suggests the presence of a quantum phase transition at $x \sim 0.5$, where nematic order disappears and a gap opens up in the spin fluctuation spectrum. While the spin fluctuations above the gap remain characteristic of the ordered phase nearby, the dc-transport properties appear to be primarily controlled by interactions with low-energy excitations.

The energy and momentum dependence of the spin excitations we have observed helps to develop a microscopic description of the nematic state, and to discriminate between different theoretical descriptions of the coupling between spin and charge excitations in the cuprates. Specifically, the data of Fig. 3A are incompatible with theories according to which the resistivity anisotropy is controlled by an anisotropic spin-spin correlation length (*30*). Rather, the data indicate a correspondence between the upturn in $\rho_a/\rho_b$ and the onset of the incommensurate modulation of the spin system. Incommensurate peaks in the magnetic neutron scattering pattern can arise either from a longitudinal modulation, where the magnetic moments are collinear but their amplitude is spatially modulated, or from a transverse modulation, where the moment direction varies but the amplitude remains constant. Our data are compatible with slow fluctuations characteristic of either type of modulation. Spin-amplitude modulated states naturally go along with a modulation of the charge carrier density, and the carrier mobilities along and perpendicular to the modulation axis are generally expected to be different (*7*). It has also been shown that a transverse modulation with spiral spin correlations can lead to anisotropic hopping transport in weakly doped cuprates with diverging low-temperature resistivity (*31*). Further work is required to assess whether this mechanism can be generalized to metallic electron systems such as the one in $YBa_2Cu_3O_{6.45}$. Finally, we note that the magnitude of $\delta$ in $YBa_2Cu_3O_{6.45}$ is incompatible (*16,18*) with the Yamada-plot (*32*), that is, the $\delta$-versus-$x$ relation that holds generally for La214: This relation stipulates an incommensurability $\delta \sim 0.085$ r.l.u. for our doping level of 0.085 holes per Cu atom, while we observe $\delta$ to saturate at $\sim 0.045$ r.l.u. at the lowest energies and temperatures. Besides, a spontaneous onset of the incommensurability has thus far not been observed in this cuprate family (*2*).

The spin dynamics of $YBa_2Cu_3O_{6.45}$ and its close correspondence with the charge transport properties of this material provide strong evidence of a cooperative transition to an electronic nematic phase at a temperature that is about two orders of magnitude higher than the onset of static magnetic order. In the light of prior work on La214 (*33*), we expect the quasi-static, incommensurate magnetic order observed here to be enhanced by a magnetic field perpendicular to the $CuO_2$ planes, with possibly important consequences for the interpretation of the Fermi surface pockets inferred from recent high-field quantum oscillation measurements on Y123 crystals with doping levels similar to ours (*34,35*).

**Acknowledgements**: Part of this work is based on experiments performed at the Swiss muon source SμS, Paul Scherrer Institute, Villigen, Switzerland. We thank A. Suchaneck, S. Pailhès, Ch. Niedermayer and A. Amato for help during the experiments. V.H., C.B. and B.K. acknowledge financial support by the Deutsche Forschungsgemeinschaft in the consortium FOR538 and C.B. by the Schweizer Nationalfonds (SNF) via grant 200021-111690/1. We thank Y. Ando, P. Hirschfeld, G. Khaliullin, S. Kivelson, D. Manske, O. Sushkov, J. Tranquada, M. Vojta, H. Yamase and R. Zeyher for helpful discussions.

## References and Notes

1. J. M. Tranquada, B. J. Sternlieb, J. D. Axe, Y. Nakamura, S. Uchida, *Nature* **375**, 561 (1995).

2. M. Fujita, H. Goka, K. Yamada, J. M. Tranquada, L. P. Regnault, *Phys. Rev. B* **70**, 104517 (2004).

3. N. B. Christensen *et al.*, *Phys. Rev. Lett.* **98**, 197003 (2007).

4. P. Abbamonte *et al.*, *Nat. Phys.* **1**, 155 (2005).

5. M. v. Zimmermann *et al.*, *Europhys. Lett.* **41**, 629 (1998).

6. S. A. Kivelson, E. Fradkin, V. J. Emery, *Nature* **393**, 550 (1998).

7. S. A. Kivelson *et al.*, *Rev. Mod. Phys.* **75**, 1201 (2003).

8. V. Cvetkovic, J. Zaanen, *Phys. Rev. Lett.* **97**, 045701 (2006).

9. V. Oganesyan, S. A. Kivelson, E. Fradkin, *Phys. Rev. B* **64**, 195109 (2001).

10. Y.-J. Kao, H.-Y. Kee, *Phys. Rev. B* **72**, 024502 (2005).

11. H. Yamase, W. Metzner, *Phys. Rev. B* **73**, 214517 (2006).

12. K. B. Cooper, M. P. Lilly, J. P. Eisenstein, L. N. Pfeiffer, K. W. West, *Phys. Rev. B* **65**,

241313 (2002).

13. R. A. Borzi *et al.*, *Science* **315**, 214 (2007).

14. Materials and Methods, as well as Supporting Text are attached at the end of this file.

15. Y. Ando, K. Segawa, S. Komiya, A. N. Lavrov, *Phys. Rev. Lett.* **88**, 137005 (2002).

16. The lattice parameters, $a$ = 3.8388 Å, $b$ = 3.8747 Å and $c$ = 11.761 Å, correspond to the chemical composition $YBa_2Cu_3O_{6.45}$ and a hole concentration of 0.085 ± 0.01 per $Cu^{2+}$ ion (*36*).

17. H. F. Fong *et al.*, *Phys. Rev. B* **61**, 14773 (2000).

18. P. Dai, H. A. Mook, R. D. Hunt, F. Dogan, *Phys. Rev. B* **63**, 054525 (2001).

19. C. Stock *et al.*, *Phys. Rev. B* **69**, 014502 (2004).

20. C. Stock *et al.*, *Phys. Rev. B* **73**, 100504(R) (2006).

21. C. Niedermayer *et. al.*, *Phys. Rev. Lett.* **80**, 3843 (1998).

22. B. J. Sternlieb *et al.*, *Phys. Rev. B* **41** 8866 (1990).

23. B. Keimer *et al.*, *Phys. Rev. B* **46**, 14034 (1992).

24. F. C. Chou *et al.*, *Phys. Rev. Lett.* **71**, 2323 (1993).

25. Y. Kohsaka *et al.*, *Science* **315**, 1380 (2007).

26. Y.-S. Lee, K. Segawa, Y. Ando, D. N. Basov, *Phys. Rev. B* **70**, 014518 (2004).

27. S. M. Hayden, H. A. Mook, P. Dai, T. G. Perring, F. Dogan, *Nature* **429**, 531 (2004).

28. V. Hinkov *et al.*, *Nature* **430**, 650 (2004).

29. V. Hinkov *et al.*, *Nat. Phys.* **3**, 780 (2007).

30. P. A. Marchetti, G. Orso, Z. B. Su, L. Yu, *Phys. Rev. B* **69**, 214514 (2004).

31. V. N. Kotov, O. P. Sushkov, *Phys. Rev. B* **72**, 184519 (2005).

32. M. Fujita *et al.*, *Phys. Rev. B* **65**, 064505 (2002).

33. B. Lake *et al.*, *Nature* **415**, 299 (2002).

34. N. Doiron-Leyraud *et al.*, *Nature* **447**, 565 (2007).

35. A. J. Millis, M. Norman, *Phys. Rev.* B **76**, 220503(R) (2007).

36. R. Liang, D. A. Bonn, W. N. Hardy, *Phys. Rev. B* **73**, 180505 (2006).

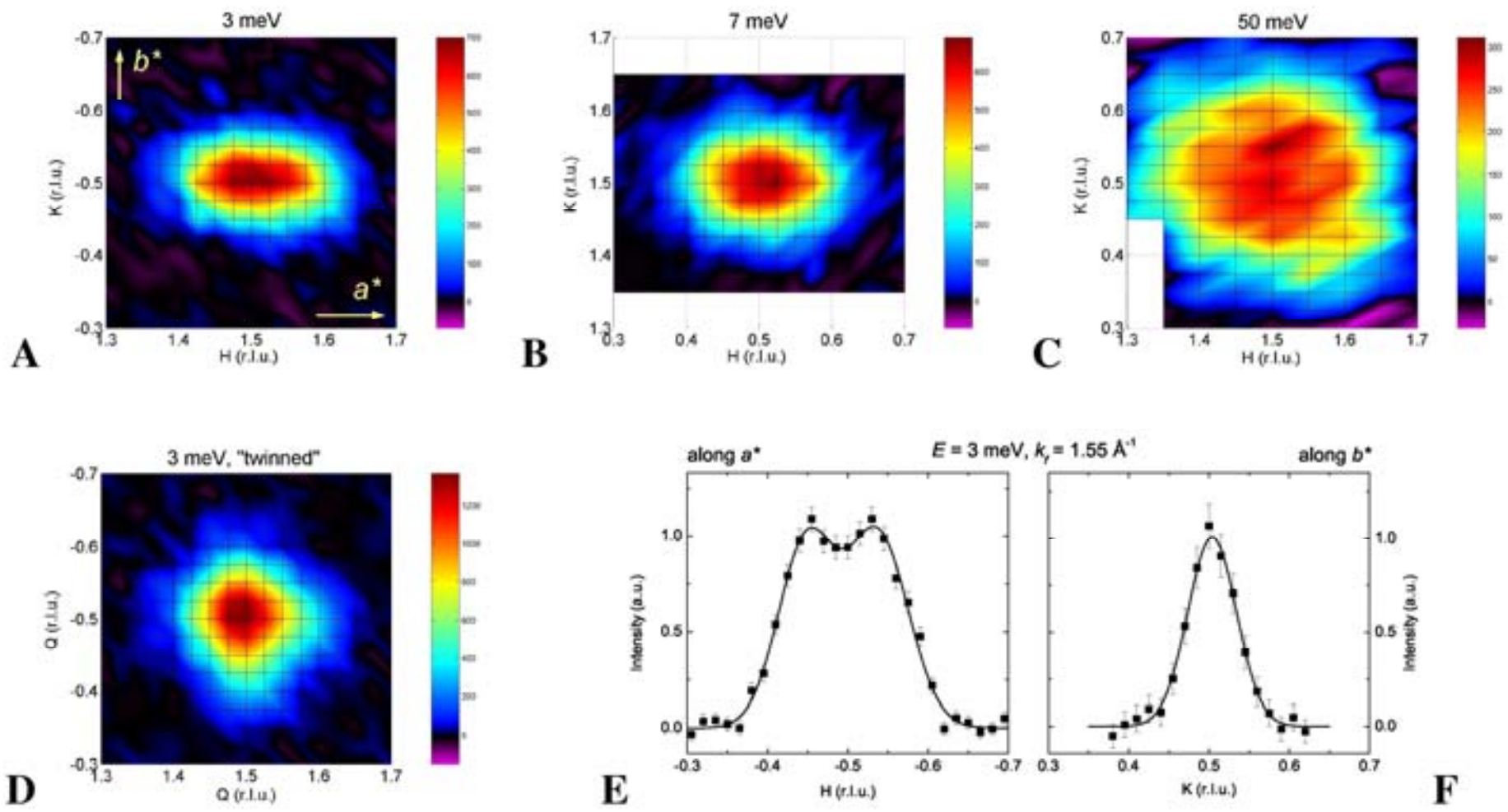

Fig. 1. Geometry of the spin excitations around $\mathbf{Q}_{\mathrm{AF}}$ at L = 1.7 r.l.u. in the *a*-*b* plane at $T = 5$ K. (**A–C**) Intensity maps of the spin-excitation spectrum at 3, 7 and 50 meV, respectively, assembled from triple-axis scans. The $a^*$-and $b^*$-directions are indicated in panel A. The scale is in arbitrary units and the wave vector is in reciprocal lattice units (r.l.u.), see also the SOM (*14*). The wave vector of the scattered neutrons, $k_f$, was fixed to 2.66 Å$^{-1}$ in A,B and to 4.1 Å$^{-1}$ in C. The crossings of black lines represent measured data points. All scans were corrected for a **Q**-linear background. (**D**) Color map of the intensity at 3 meV, as it would be observed in a crystal consisting of two perpendicular twin domains with equal population. The representation was obtained by transposing the map in A and superposing it with the original map. (**E,F**) Scans along $a^*$ and $b^*$ through $\mathbf{Q}_{\mathrm{AF}}$. The resolution was enhanced as compared to A–D by working with $k_f = 1.55$ Å$^{-1}$. Solid lines represent fits with one (F) or two (E) Gaussians to the data. A linear background was subtracted.

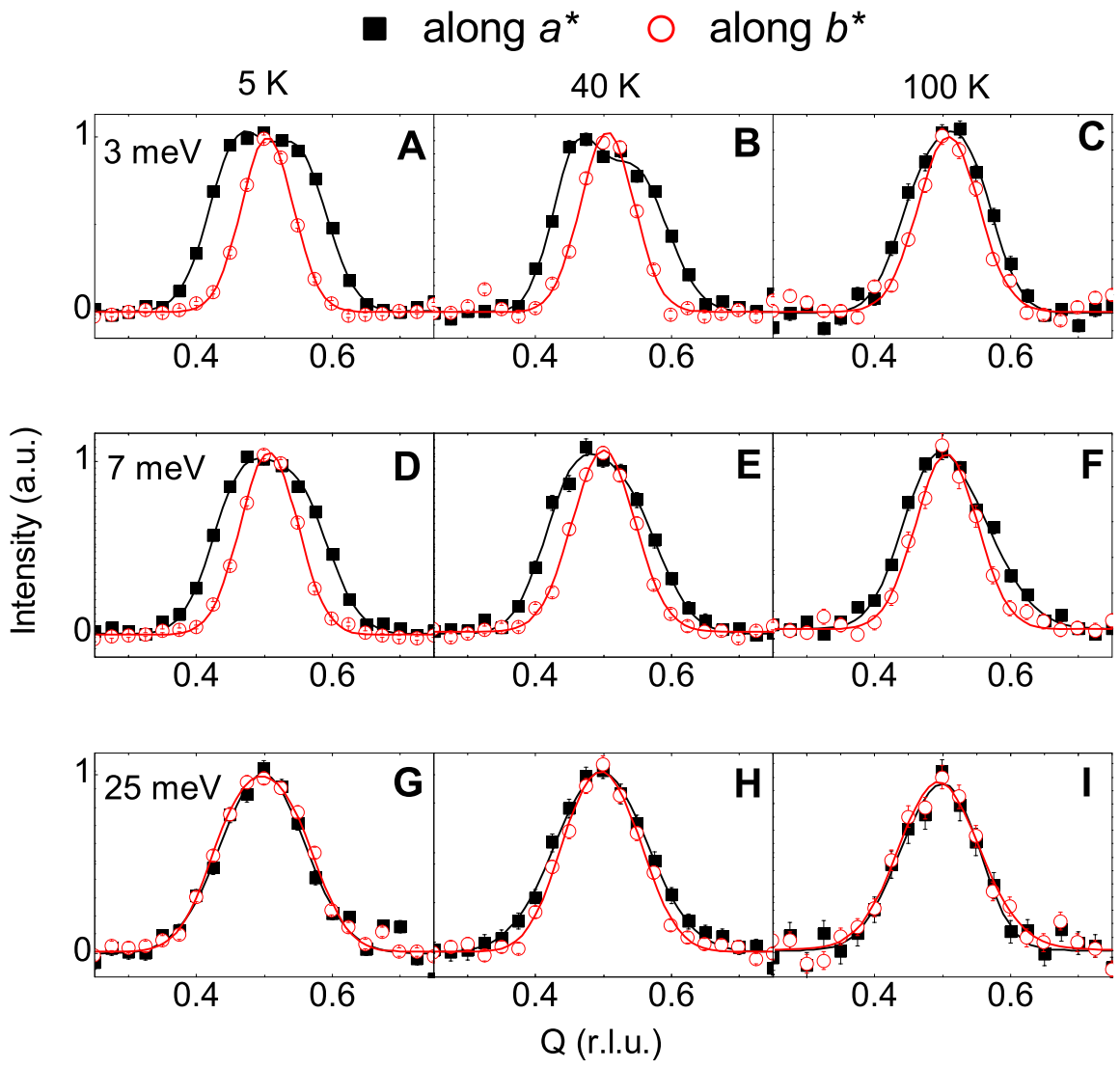

Fig. 2. Temperature evolution of the *a*-*b* anisotropy of the spin correlations. Full squares and empty circles represent data points measured at fixed $K$ along $a^*$ and at fixed $H$ along $b^*$, respectively. (**A–C**) $E = 3$ meV, (**D–F**) $E = 7$ meV, (**G–I**) $E = 25$ meV. The measurements were performed at $T = 5$ K (**A, D, G**), $T = 40$ K (**B, E, H**) and $T = 100$ K (**C, F, I**). The final wave vector $k_f$ was fixed to 2.66 . Å$^{-1}$. All scans are normalized to unity in order to allow a better comparison of the scan profiles. Solid lines represent the results of fits with one or two Gaussians. A linear background was subtracted.

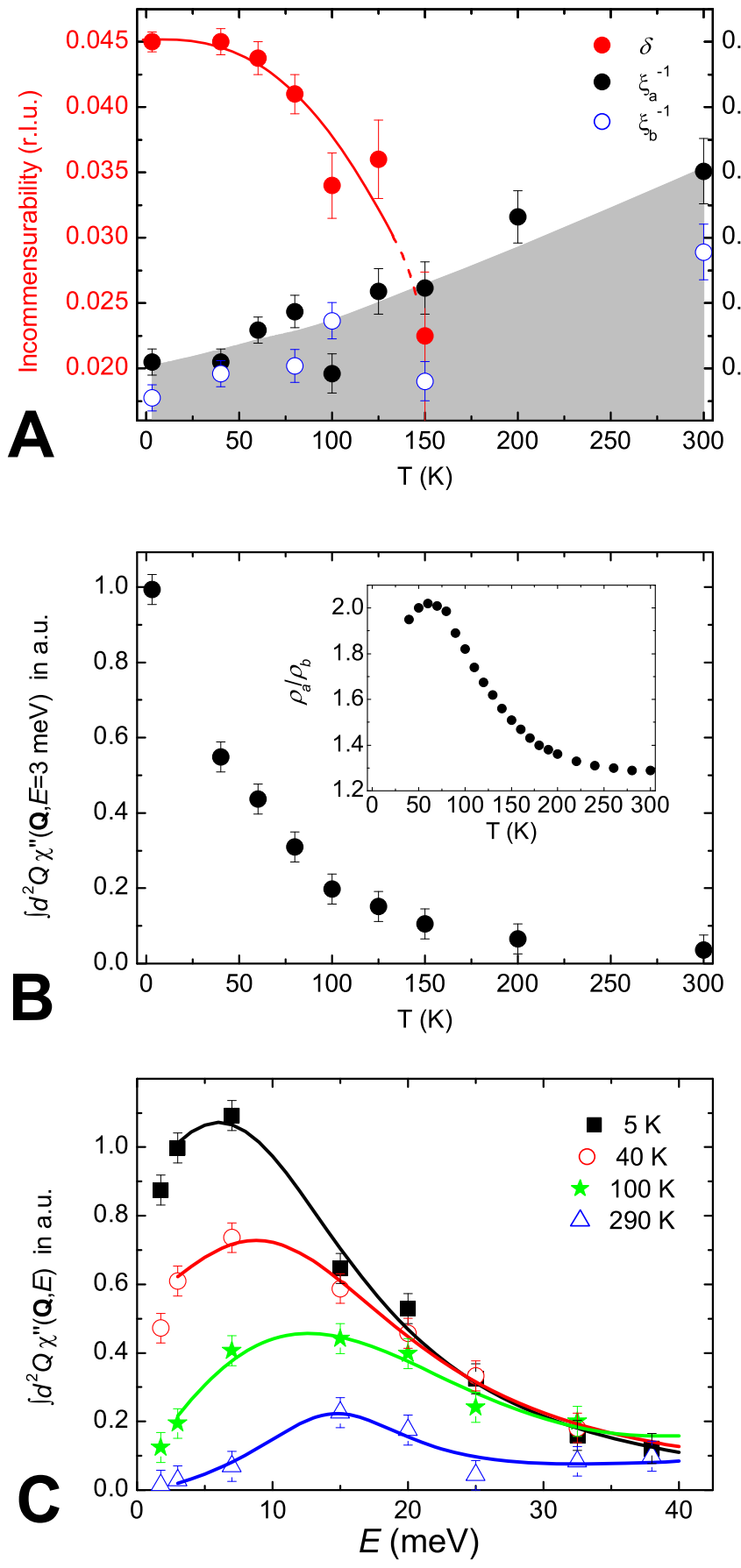

Fig. 3. Temperature and energy evolution of parameters characterizing the spin excitation spectrum. The parameters are the results of fits to the raw data, corrected for the instrumental resolution. (**A**) Incommensurability $\delta$ (red symbols), half-width-at-half-maximum of the incommensurate peaks along $a^*$ ($\xi_a^{-1}$, black symbols) and along $b^*$ ($\xi_b^{-1}$, open blue symbols) in reciprocal lattice units. The upper border of the shaded area follows $\xi_a^{-1}$. The $\xi^{-1}$-axis is scaled to twice the value of the $\delta$-axis, hence $\delta$-points lying inside the shaded area indicate an IC peak separation below $\xi_a^{-1}$. (**B**) Imaginary part of the **Q**-integrated spin susceptibility $\chi''(\omega)$ at 3 meV. The inset shows the ratio of the electrical resistivity along $a^*$ and $b^*$ of a sample with similar doping level as ours, reproduced from Ref.(*15*). (**C**) Energy evolution of $\chi''(\omega)$ at $T = 5$, 40, 100, and 290 K. In all panels, the error bars were estimated from the fits. The lines are guide to the eye.

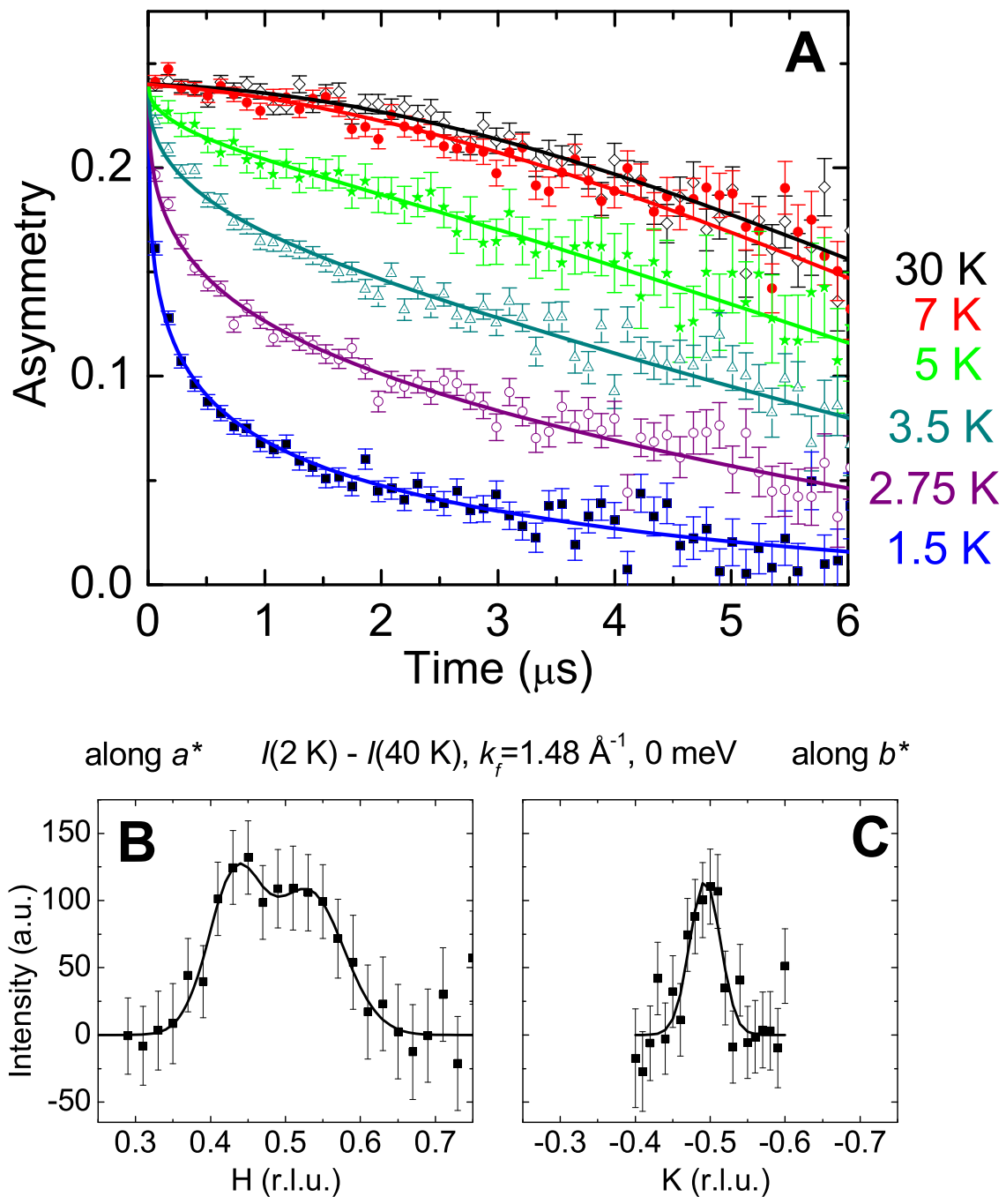

Fig. 4. Measurements of the quasi-elastic magnetic response. (**A**) Zero field muon-spin-relaxation data taken at $T = 1.5$, 2.75, 3.5, 5, 7, and 30 K. Solid lines show the results of fits to a relaxation function that consists of the product of a stretched exponential and a Kubo-Toyabe function, which account for the contribution of the electronic and nuclear magnetic moments, respectively. (**B,C**) Neutron scattering scans along $a^*$ at fixed $K = -0.5$ and along $b^*$ at fixed $H = 0.54$, with a nominal energy transfer of 0 meV. The difference between the intensities at 5 K and 40 K is shown in arbitrary units and $k_f$ was fixed to 1.48 Å$^{-1}$. Solid lines represent fits with one (C) or two (B) Gaussians to the data. A linear background was subtracted.

# Materials and Methods

## Sample preparation

As-grown $YBa_2Cu_3O_{6+x}$ samples are twinned, i.e. they consist of equal proportions of twin domains with mutually perpendicular orientation of the in-plane axes *a* and *b*. To probe the in-plane anisotropy of the spin correlations, a large, almost twin-free (domain population ratio > 92%) sample array with a mosaicity ~ 1° was prepared consisting of ~ 100 crystals with a total mass of ~ 2 g. All crystals were characterized by magnetometry and were found to exhibit superconducting transition temperatures Tc = 35 K with transition widths of 2–3 K, testifying to their high quality. The lattice parameters, *a* = 3.8388 Å, *b* = 3.8747 Å, and *c* = 11.761 Å, correspond to the chemical composition $YBa_2Cu_3O_{6.45}$ and a hole concentration of 0.085 ± 0.01 per copper ion (S1), in agreement with the superconducting transition temperature.

For the neutron scattering experiments the crystals were arranged on a thin aluminum plate, see fig. S1. Each crystal was aligned individually, using an X-ray Laue device. The resulting crystal array has a mosaicity of ~ 1°.

## Neutron scattering measurements

The neutron measurements were carried out at the triple-axis spectrometers IN8 and IN14 (Institut Laue Langevin) and 4F2 (Laboratoire Léon Brillouin). Pyrolytic graphite monochromators and analyzers were used. As described previously (S2,S3), scans along $a^*$ and $b^*$ were carried out under identical instrumental resolution conditions. No collimators were used, and graphite or beryllium filters extinguished higher-order contamination of the neutron beam. The magnetic origin of the quasi-elastic signal was confirmed in a measurement with polarized neutrons. We quote the wave vector $\mathbf{Q} = (H,K,L)$ in reciprocal lattice units (r. l. u.) with primitive vectors $a^*$, $b^*$, and $c^*$, where $a^* = 2\pi/a$, $b^* = 2\pi/b$ and $c^* = 2\pi/c$.

## Muon-spin rotation μSR measurements

The muon-spin rotation (μSR) measurements were carried out with the GPS setup at the πm3 beamline at the Paul-Scherrer-Institut in Villigen, Switzerland, which provides 100% spin-polarized muons.

# Supporting Text

## Bilayer modulation

The Y123 structure consists of $CuO_2$ bilayers separated by a layer containing CuO chains. Magnetic interactions within a bilayer unit split the magnetic excitations into branches that are odd and even under the exchange of the two layers. The excitations exhibit an *L*-dependent structure factor, and their intensity is modulated by $\sin^2(\pi Ld/c)$ and $\cos^2(\pi Ld/c)$, respectively, where d = 3.285 Å is the intra-bilayer spacing. In this work we show odd excitations only and work at the first maximum of their structure factor, 1.7 r.l.u. Even excitations show a gap of about 50 meV and do not interfere with our data (S4). In our measurements the presence of the $\sin^2$ intensity modulation was confirmed. In particular, the magnetic intensity is zero at $L = 0$ within the experimental error, see Fig. S2. This implies that magnetic moments in the CuO chains do not contribute to the spin correlations shown here, and that the signal shown derives entirely from the $CuO_2$ planes.

## Figures

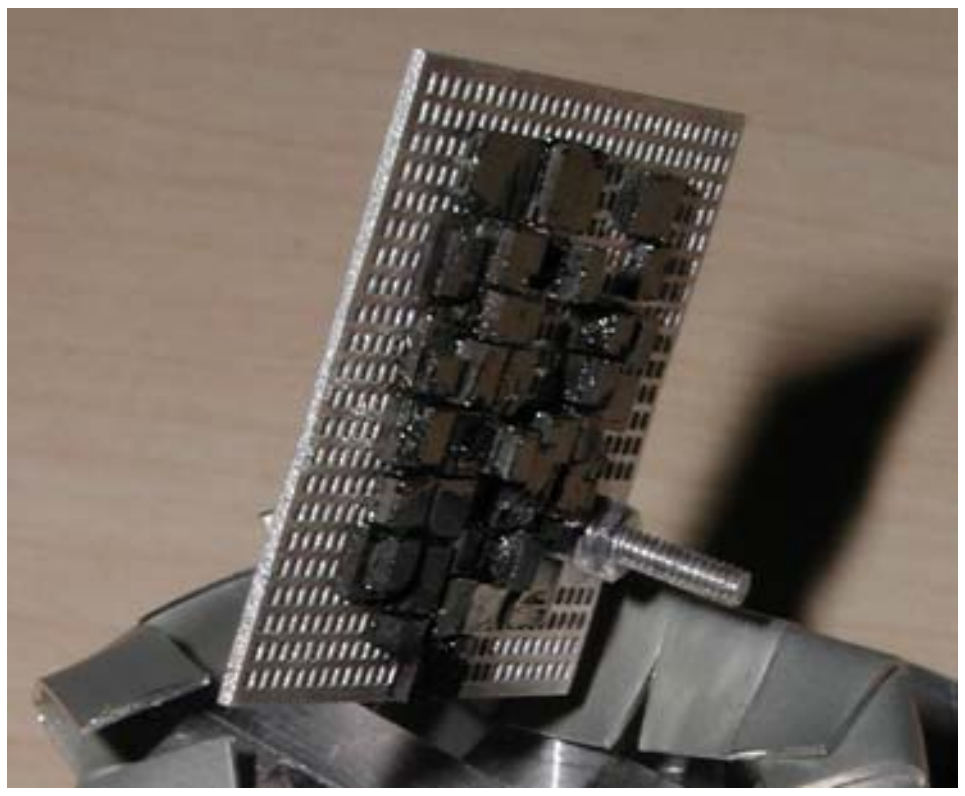

Fig. S1: Photograph of the $YBa_2Cu_3O_{6.45}$ sample used in this experiment. The Al-plate carrying the crystals is perforated to reduce the amount of Al in the neutron beam, which can give rise to spurious scattering. The *c*-axis is perpendicular to the plate surface. Up to three crystals are stacked on top of each other (the back side of the plate is similarly covered with crystals). This is to concentrate the sample in the middle of the beam and to avoid the necessity of a second Al-plate.

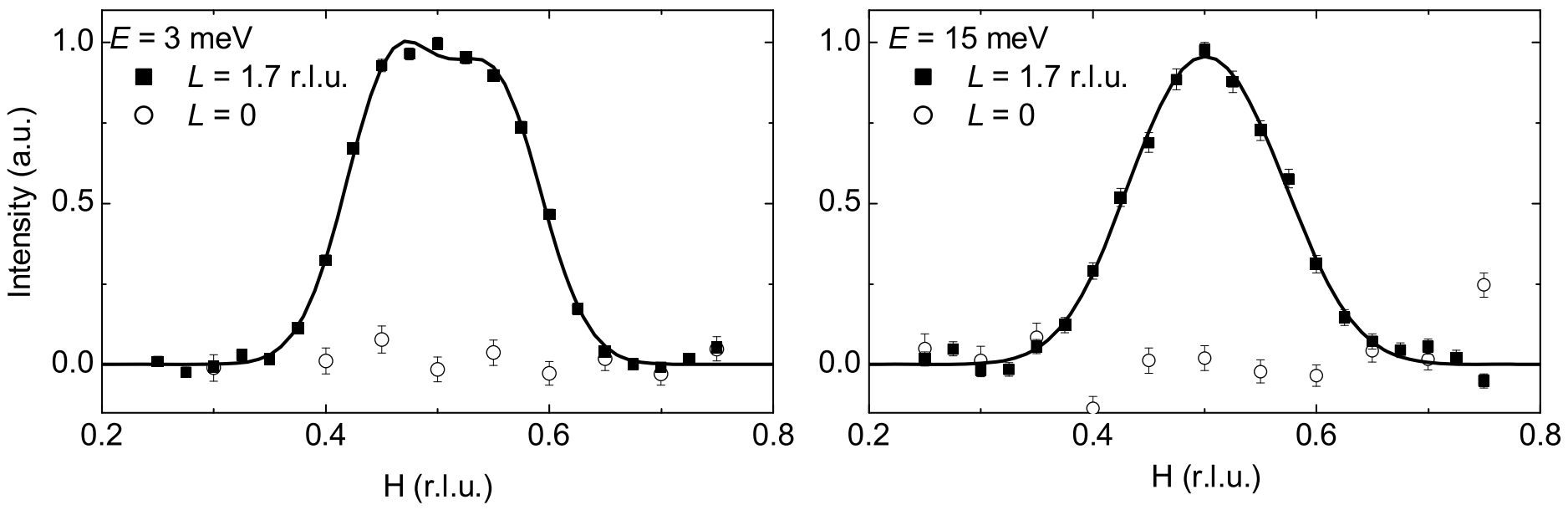

Fig. S2: Scans along *a** through $\mathbf{Q}_{AF}$ at 3 meV (left panel) and 15 meV (right panel) at 5 K. Black symbols show the intensity at the maximum of the odd bilayer structure factor ($L = 1.7$ r.l.u.), while open symbols show the intensity at its minimum ($L = 0$). A linear background has been subtracted.

## References

S1. R. Liang, D. A. Bonn, W. N. Hardy, Phys. Rev. B 73, 180505 (2006).
S2. V. Hinkov et al., Nature 430, 650 (2004).
S3. V. Hinkov et al., Nat. Phys. 3, 780 (2007).
S4. H. F. Fong et al., Phys. Rev. B 61, 14773 (2000).